\documentclass[prl,aps,twocolumn,groupedaddress,floats,showpacs,final]{revtex4-1}
\usepackage{graphicx}
\usepackage{dcolumn}
\usepackage{bm}
\usepackage{color}
\definecolor{blue}{rgb}{0.3,0.3,0.9}
\def\he4{$^4$He}
\def\hee3{$^3$He}
\def\beq{\begin{eqnarray}}
\def\eeq{\end{eqnarray}}

\begin{document}

\title{Emergence of Luttinger Liquid behavior of a superclimbing dislocation }

\author{M. Yarmolinsky}
\author{A. B. Kuklov}
\affiliation {Department of Engineering \& Physics and the Graduate Center, CUNY, Staten Island, NY 10314, USA}

\date{\today}
\begin{abstract}
A superclimbing dislocation spread over several valleys of Peierls potential in solid \he4  represents a non-Luttinger Liquid according to the elementary scaling dimensional analysis because its excitation spectrum is parabolic. Monte Carlo simulations, however, reveal that such a dislocation develops  Luttinger Liquid behavior, which can also undergo a transition into insulating state.
 External bias can restore the parabolic spectrum. An experimental verification of the effect is proposed.

\end{abstract}

\maketitle

Emergence \cite{emergence} of unexpected behavior is the topic drawing a lot of attention for the last 50 years.  
The most prominent example is the charge fractionalization in fractional quantum Hall effect \cite{FQH}. Dynamical enlargement of Hamiltonian symmetry group at the point of continuous phase transition \cite{scaling} represent yet another playground where the emergence takes place. 

Here we introduce a system which demonstrates emergence of Luttinger Liquid (LL) from a non-LL in solid \he4. It is the so called  superclimbing dislocation which is tilted in the Peierls crystal potential. The concept of superclimb has been proposed in Ref.\cite{sclimb} as a possible explanation for the giant isochoric compressibility (syringe effect) serendipitously observed in the  UMASS group \cite{Hallock} during the superflow through solid \he4 events. In its essence, a solid exhibits the response on external chemical potential applied at a point, practically, the same way as  liquid does --  absorbs or expels a macroscopic fraction of atoms.      
This effect has also  been seen by the Univ. of Alberta group, \cite{Beamish}, and very recently confirmed in its most conspicuous form  in Ref.\cite{Beamish_2016}.

 Based on the {\it ab initio} simulations \cite{sclimb} revealing superfluid core of edge dislocation with Burgers vector along the {\it hcp} axis,  it has been suggested that the syringe effect  can be associated with the so called superclimb -- climb assisted by the superfluid transport along dislocation core. The key feature of this scenario is that the isochoric compressibility (response on chemical potential $\mu$ at finite volume) of a solid permeated by a uniform network of dislocations with superfluid core is independent of  density of the superclimbing dislocations and is, instead, determined by the dimensionless parameter -- the asymmetry between lengths of superclimbing and non-superclimbing parts (screw dislocations \cite{screw}). This implies that the effect is strongly non-perturbative, with respect to dislocation density. 

The model \cite{sclimb}  features a dislocation subjected to Peierls potential and interacting with its own superfluid core through the Berry term accounting for the mass transferred through the core which contributes to introducing/removing crystalline planes of the solid. The elementary scaling analysis shows, Ref.\cite{sclimb}, that the Peierls potential is always relevant at zero temperature, $T=0$, so that the dislocation aligned with a Peierls valley must be "pinned" to the valley and therefore it cannot demonstrate superclimb. Such a self-pinning guarantees that the core superfluid behaves as LL (that is, characterized by linear dispersion of excitations \cite{Haldane}). As temperature increases the climb becomes possible. This results in the compressibility $\kappa$ becoming "giant" as $\kappa \propto L^2$ where $L$ is a typical length of a free segment of the dislocation. Accordingly, the excitation spectrum becomes parabolic.    

If the dislocation is not aligned with one Peierls valley, the Peierls potential becomes irrelevant even at $T=0$. 
In this case jogs exist even at $T=0$ because their number is determined by the number of the valleys hosting the dislocation.
 Since such jogs are quantum objects, their quantum fluctuations wash out the Peierls potential \cite{PRB2014} and therefore the dislocation should remain in the superclimb regime even at $T=0$ -- at least at the gaussian level of the model \cite{sclimb}.   This simple argument, however, has not been verified numerically.

Here we analyze a tilted superclimbing dislocation beyond the guassian approximation by Monte Carlo simulations of the model \cite{sclimb} where Peierls potential is eliminated. Our main result is that there is a crossover from non-LL to LL as temperature $T$
and $L$ both scale as $T \sim 1/L \to 0$. An external bias of the dislocation by chemical potential $\mu$ can destroy the LL and restore the superclimb as long as it exceeds a threshold which is {\it macroscopically} small with respect to $L$.

\noindent {\it The model}.
In its gaussian form the action 
\beq
S&=& \int_0^\beta d\tau \int_0^L dx [-  i(y +n_0)\partial_t \phi  + \frac{\rho_0}{2} (\partial_x\phi)^2  
\nonumber \\
&+&\frac{\kappa_0}{2} (\partial_t \phi)^2 + {G\over{2}}(\partial_x y)^2   
 - \mu y(x,t)] ,
\label{S}
\eeq
(in units $\hbar =1, K_B=1$) depends on the displacement of the dislocation $y(x,\tau)$ from its equilibrium position and on the
superfluid phase $\phi(x,\tau)$ along the superfluid core. Here $\beta =1/T$, $\rho_0$ and $\kappa_0$ are bare superfluid stiffness and superfluid copmressibility, respectively; The quantity $n_0$ describes average filling factors; $G$ stands for the effective tension of the dislocation ($ \sim $ shear modulus). 

The imaginary term in Eq.(\ref{S}) is the Berry term counting how many particles passed through the dislocation core and ended up in an extra row of atoms advancing dislocation by $y$. In other words,
if $y$ changed by $\delta y$ over the whole dislocation length $L$, the extra matter delivered (or removed) to the solid is $ L \delta y$. 
In the action (\ref{S}) we are interested in the large wavelengths  and small frequencies limit. Thus, the time derivative
$ \sim (\partial_\tau y)^2$ representing kinetic energy of the dislocation motion is omitted because the flow along the core controls the dynamics.   In order to exclude the zero mode
where the uniform shift of the dislocation as a whole costs no energy, we will be considering the boundary condition $y(x=0,\tau)=
y(x=L,\tau)=0$.

Variational equations of motion $\delta S/\delta \phi=0, \, \delta S/\delta y=0 $ following from the action (\ref{S}) in the gaussian approximation are
$ \partial^2_{\tau} y - G\rho_0 \partial^4_x y =0$ in the long wave limit. In real time $t=i \tau$ it corresponds to the spectrum 
$\omega = \sqrt{G\rho_0} q^2$, which is distinctly different from the standard LL one $\omega=V_s q $ with the speed of sound $V_s =\sqrt{\rho_0/\kappa_0}$.   

Here we will go beyond the gaussian approximation by taking into account the compact nature of the phase $\phi$ by allowing 
vortices (instantons) to exist in the space-time $(x,\tau)$. This, in particular, can be achieved by discretizing the space-time so that $\int d\tau \int dx ...$ transforms into a sum over the space-time lattice. Then, the continuous derivatives become discrete:
$\partial_x \phi(x,\tau) \to \nabla_x \phi = \phi(x+1,\tau) - \phi(x,\tau),\, \partial_\tau \phi(x,\tau) \to \nabla_\tau \phi =[ \phi(x,\tau +\Delta \tau) - \phi(x,\tau)]/\Delta \tau$, where the increment of space is taken as unity and $\Delta \tau$ is the unit of the time discretization $\Delta \tau = \beta/N_t \to 0$, with $N_t$ being the number of time slices in the time interval $(0,\beta)$. 
 Then, the compactness of $\phi$ is taken into account by using the Villain \cite{Villain} approximation $   \vec{\nabla} \phi \to \vec{\nabla} \phi + 2\pi \vec{m}$, where the vector sign refers to the space-time directions and $\vec{m}$ stands for integers variables defined on bonds between neighboring sites of the
space-time lattice. Then, $\phi$ can be treated as a non-compact gaussian variable on the expense of introducing the bond variables $\vec{m}$. 

The thermodynamics of the model (\ref{S}) (with the substitute $   \vec{\nabla} \phi \to \vec{\nabla} \phi + 2\pi \vec{m}$ can be accounted for within the partition function    
\beq
Z= \sum_{\{\vec{m}\}} \int D\phi \int Dy \exp(-S).
\label{Z}
\eeq
 It is convenient to use Poisson identity
$\sum_m f(m) \equiv \sum_n \int dm f(m) \exp(2\pi imn)$ at each bond along the line of the derivation of the J-current model \cite{Jcur}. Then, the integrations over $\vec{m}, \, \phi, \, y$ can be carried over exactly. This transforms Eq.(\ref{Z}) into
$Z= \sum_{\{\vec{J}=(J_x,J_\tau)\}} \exp(-S_J)$,
where the action $S_J$ in the long-wave limit is
\beq
S_J= \sum_{b_{ij}}\left[\frac{J_x^2}{2\tilde{\rho}_0} + \frac{\tilde{G}}{2} (\nabla_x J_\tau)^2 - \tilde{\mu} J_\tau\right], 
\label{S_J}
\eeq
with $\tilde{G}=G\Delta \tau$, $\tilde{\mu}=\mu \Delta \tau$ and $\tilde{\rho}_0=1/[2 \ln(2/\rho_0\Delta \tau)]$ (in the limit $\Delta \tau \to 0$ \cite{Villain}). 
The integer bond oriented currents $\vec{J}$ (between neighboring sites) satisfy the Kirchhoff's conservation rule,
and the summation is performed over all bonds $b_{ij}$ between all pairs of neighboring
 sites $i$ and $j$. It should be kept in mind that $\vec{J}=(J_x,J_\tau)$ is oriented either along a spatial or a temporal bond. 
In other words if $b_{ij}$ is a bond along X-direction, the current along this bond has zero temporal component, $J_\tau=0$.
Similarly,  $J_x=0$ on a bond oriented along the imaginary time axis.

In the action (\ref{S_J}) only the lowest gradient $ \nabla_x J_\tau \equiv J_\tau(x+1,\tau) - J_\tau(x,\tau)$ has been considered in the low energy limit.
The boundary condition for $y$ is transformed into $J_\tau(x=0,\tau)=J_\tau(x=L,\tau)=0$ in addition to the periodic boundary
conditions. 

The striking difference between the action (\ref{S_J}) and the standard one of the J-current model \cite{Jcur} is the absence of the
term $\sim J_t^2$. As we will show below such a term will be {\it emerging} as $T\to 0$ and $\mu \to 0$. 

\noindent{\it Linear response}.
The linear response  of the system is described in terms of the renormalized stiffnesses along space and along time \cite{Polock}. The first one represents the renormalized superfluid stiffness $ \rho_s = LT \langle W_x^2\rangle, \,\, W_x=\frac{1}{L} \sum_{\vec{b}_{ij}} J_x$,
and the second is the renormalized compressibility 
$
\kappa= - \frac{T}{L} \frac{\partial^2 \ln Z}{\partial \mu^2}=\frac{1}{LT} \langle W^2_\tau \rangle , \,\, W_\tau=T  \sum_{b_{ij}} J_\tau $.
The quantities $W_x$, $W_\tau $ are integers and have the geometrical meaning -- windings of the lines formed by the J-currents.
These responses  have been  evaluated numerically by the Worm Algorithm \cite{WA}.

\noindent{\it Giant isochoric compressibility.}
As suggested in Ref.~\cite{PRB2014}, the action (\ref{S}) should describe superclimb because there is no Peierls potential.
The compressibility in this regime can be evaluated in the gaussian approximation as
\beq
\kappa=\kappa_g= \frac{L^2}{12G}.
\label{Kap_0}
\eeq
It is important to note that the factor $L^2$ is canceled from the 3D compressibility of the solid permeated by the dislocation network. Indeed, let's presume the network is characterized by a typical length $L$ of the superclimbing segments. Then,
each segment accepts $\delta N = \kappa_g L \delta \mu$ of extra particles in response to a small change $\delta \mu$ of $\mu$. 
Thus, each elementary cube with side $\sim L$ of the network gains $ \sim 3 \delta N \sim L^3\delta \mu/(4G) $ particles. Accordingly, the solid density changes by $ \sim \delta \mu/(4G)$ which is independent of $L$. Such a response is essentially that of a liquid. Clearly, applying $\delta \mu$ to ideal solid at a point contact does not change density by any detectable amount.     

It should also be mentioned that Eq.(\ref{Kap_0}) is valid as long as $\mu$ is below the value $ \mu_c \approx G/(2\pi L)$
 which determines the threshold for the superclimbing instability leading to the unrestricted growth of the dislocation \cite{PRB2015}.
Eventually such a dislocation can exit the solid and leave behind one extra layer of atoms. This process
represents a mechanism of crystal growth from inside out -- as an alternative to the growth from the surface.
The superclimb can be also stopped by accumulating stress in the solid due finite density of dislocations \cite{PRB2015}.

\begin{figure}[t]
\centerline{\includegraphics[width=1.1\columnwidth, angle=0]{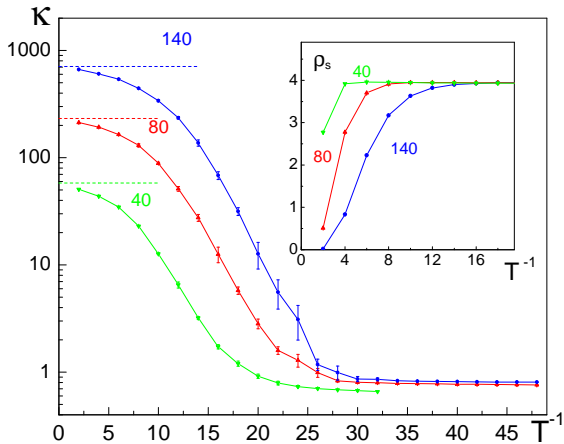}}
\vspace*{-0.5cm}
\caption{(Color online)  $\kappa$ vs $\beta=T^{-1}$ for three values of $L=40,80,140$ (shown close to each curve). The horizontal dashed lines are the corresponding values of the "giant" compressibility, Eq.(\ref{Kap_0}). Inset: superfluid stiffness vs $T^{-1}$ for the same sizes. The model parameters are $\rho_0=4, G=2.3, \mu=0$ in Eq.(\ref{S_J}).}
\label{Fig:Kap}
\end{figure}

\noindent{\it Emergence of the LL}.
Strictly speaking, all results of simulations of the model (\ref{S_J}) should be considered in the limit $N_\tau \to \infty$ in order to achieve the limit of continuous time. Practically, $N_\tau$ should be taken as large as needed to stop simulated quantities being dependent on $N_\tau$ for a given value of $\beta$.
We have checked that, while changing specific values, the qualitative behavior remains the same for a fixed value of $N_\tau$ 
without formally achieving the quantum limit of continuous time. Thus, the discussion below will be presented for $T=1/N_\tau$, that is, for the choice $\Delta \tau =1$ (and  $\tilde{G}=G, \tilde{\rho}_0=\rho_0, \tilde{\mu}=\mu$).
The results of the simulations of the model (\ref{S_J}) are presented in Fig.~\ref{Fig:Kap}.
As can be seen, with no  bias, $\mu=0$, the compressibility decreases from the "giant" values (\ref{Kap_0}) to some finite one upon decreasing temperature.  
\begin{figure}[t]
\centerline{\includegraphics[width=1.1\columnwidth, angle=0]{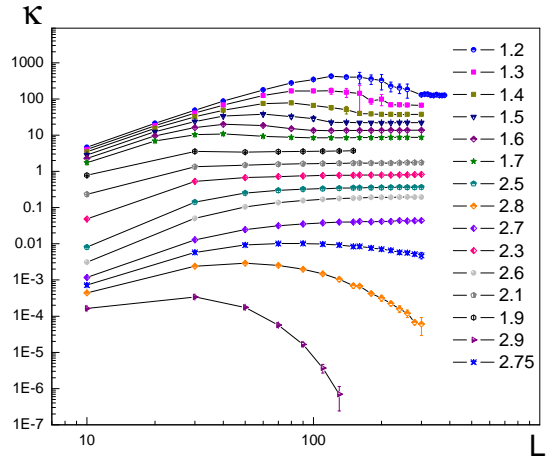}}
\vspace*{-0.5cm}
\caption{(Color online)  $\kappa$ vs $L=1/T$ for various values of $G$ and $\rho_0=4, \mu=0$. }
\label{Fig:KapG}
\end{figure}
In order to evaluate the crossover temperature $T_L$ and its width $\Delta_L$ we have found the best fit of $\kappa$ vs $1/T$ using $T_L$ and $\Delta_L$ as the fit parameter in the fit function taken as $ \ln(\kappa)= A -B \cdot \tanh(\Delta_L\cdot (T^{-1} - T^{-1}_L))$,
with $A$ and $B$ chosen from the limiting values of $\kappa$ at the highest and lowest $T$ for each $L$. This function has produced  fits which are acceptable within the statistical errors of the data for all curves. We have found that, $T_L \sim 1/\ln L$ and $\Delta_L \sim 1/\ln L$. More specifically, for  $G=2.3,\, \rho_0=4, \, \mu=0$, the dependencies on $L$ are $T^{-1}_L = a \ln L +b$, with $a=5.02, \, b= - 6.27$ and $\Delta^{-1}_L=a \ln L +b$ with $  a=1.53,\, b=0.09$.   

The question is how the emerged compressibility in the limit $L=\infty$ depends on the parameters of the model (\ref{S_J}). Fig.~\ref{Fig:KapG} presents results of simulations for $\rho_0=4$ and $1/T=L$ for various values of $G$ shown in the legend. The limiting value of $\kappa_{eff}=\kappa$ taken from the saturated behavior at large $L$ from Fig.~\ref{Fig:KapG}   turns out to
be proprtional to high negative power of the paramer $G$  in the action (\ref{S_J}) as $\sim 1/G^b,\, b=7.8 \pm 0.1$ for $G<2.6$. In other words, the giant compressibility $ \sim L^2/G$ as a function of $L$ levels off at some size $L^*$ such that $(L^*)^2/G \sim 1/G^b$, that is,  $L^* \sim 1/G^{3.4}$.  We have tested several values of $\rho_0$ and didn't find any dependence of the power $b$ on it. 

\noindent{\it Quantum phase transition.}
As can be seen in Fig.~\ref{Fig:KapG}, there is a value of $G\approx 2.7$ above which  finite compressibility vanishes in the limit $L \to \infty$. This means that the system flows to the insulating phase. 
The presence of the transition in the model (\ref{S_J})  is unexpected because the Kosterlitz-Thouless (KT) argument for vortex proliferation indicates that there should be no such a proliferation. Let's demonstrate this by performing duality transformation on the model (\ref{S_J}). The Kirchhoff's constraint on the currents $\vec{\nabla} \vec{J}=0 $ can be satisfied by the substitute $
J_x=\nabla_\tau \Phi,\, J_\tau= -\nabla_x \Phi$, 
where $\Phi $ are integers defined at sites of the dual lattice constructed on sites in the middle of plaquettes of the original lattice \cite{Polyakov}.
 Using this  in Eq.(\ref{S_J}) and utilizing the Poisson  summation  (along the same line how the action (\ref{S_J}) was obtained from the original one (\ref{S})), we obtain the lattice gas model $
Z=\sum_{\{n_i\}} {\rm e}^{-S_g},\, S_g=\frac{1}{2} \sum_{\vec{r},\vec{r}'} U(\vec{r}-\vec{r}')n(\vec{r})n(\vec{r}')) $,
where $n(\vec{r})$ are integers defined on the sites of the dual lattice and $U$ is the interaction potential with Fourier
components $\tilde{U}= (2\pi)^2/[ \rho^{-1}_0 \omega^2 + G q^4]$ in the low energy $\omega$ and momentum $q$ limit.
The integers $n$ describe vortices. In contrast with the standard superfluid, where vortices interact by logarithmic potential, here the potential is much stronger than logarithm. It is also strongly asymmetric: along space it is increasing with separation between two points $(x,\tau)$ and $(x,\tau')$  as $ \sim |x - x'|$ and along time  as $\sim \sqrt{|\tau - \tau'|}$. Thus, according to the    KT argument a pair of vortices with opposite vorticities cannot proliferate and destroy the fluid. However, in spite this criterion, our simulations of the model (\ref{S_J}) show that there is a transition --  the compressibility and superfluid stiffness vanish for $G> G_c$ with $G_c\approx 2.7 \pm 0.1$ in the limit $\beta = L \to \infty$. In other words, the emerged LL undergoes QPT to an insulating state. As a more detailed analysis shows, the transition corresponds to the Berezinskii-Kosterlitz-Thouless (BKT) transition with the universal jump $2/\pi$ in the effective Luttinger parameter $K=\sqrt{\rho_s \kappa}$.  It is also important to notice that the transition is insensitive to the filling factor $n_0$ in the model (\ref{S}).

 \noindent{\it Roughening transition induced by the bias $\mu$}.
The resulting LL state corresponds to smooth dislocation.
\begin{figure}[t]
\centerline{\includegraphics[width=1.1\columnwidth, angle=0]{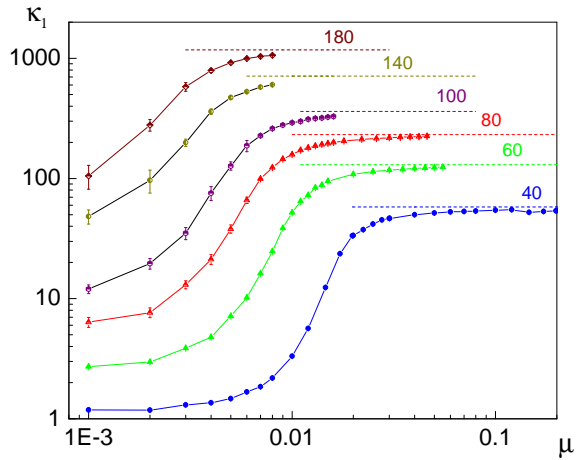}}
\vspace*{-0.5cm}
\caption{(Color online) $\kappa_1$ vs $\mu$ for sizes $L=40,60.80,100.140,180$ (shown close to the corresponding plot) , $G=2.3, \rho_0=4, 1/T=18,\, 1/T_H \approx 23$. Dashed lines show the "giant" values (\ref{Kap_0})  for the corresponding size $L$. }
\label{Fig:cross}
\end{figure}
Simulations of the model (\ref{S_J}) at finite $\mu$ have revealed two regimes: i) a crossover from smooth to rough dislocation at $T>T_H$; ii) A jump-like behavior characterized by strong hysteresis at $T<T_H$ featuring smooth and rough dislocation.
We didn't observe any consistent dependence of the special temperature $T_H$ on $L$ for sizes $L>40$. Further studies of the hysteretic behavior is definitely required.  Here we will report only on the regime i).   

The crossover behavior is shown in Fig.~\ref{Fig:cross}. 
The calculated quantity $ \kappa_1= \langle N \rangle/(L\mu)$ represents the total number of atoms $\langle N\rangle$ injected into solid due to the superclimb $N = T \sum_{b_{ij}} J_t= W_\tau$. 
This quantity coincides with $\kappa$  in the limit $\mu \to 0$.

 As can be seen, the width of the crossover becomes smaller for larger $L$. 
To characterize this dependence, we have measured the value $\mu_{0.5}$ of $\mu$ where $\kappa_1$ reaches 1/2 of its "giant" value (\ref{Kap_0}) for a given size $L$. This dependence turns out to be $ \mu_{0.5} \sim L^{-c},\, c=1.21 \pm 0.05$ for the simulated sizes $40\leq L \leq 180$ (at smaller $L$ there are strong deviations from the power law).

\noindent{\it Discussion}.
The key question to answer is why the emergence of the LL is not "seen" by the elementary dimensional analysis and also by the KT argument.
The qualitative explanation \cite{Borya}  comes naturally  in terms of the loops in Eq.(\ref{S_J}). As weight of each element $\vec{J}$ becomes larger, its discretness becomes more and more important so that more configurations will have currents $J_\tau$ with no neighbors.
In such a situation the discrete gradient $(\nabla_x J_\tau)^2 $ becomes essentially $J_\tau^2$. 

A generic network of superclimbing dislocations which may be responsible for the syringe effect \cite{Hallock,Beamish_2016} should mostly consist of the tilted dislocations. Our main prediction is that the syringe effect should vanish in the limit $T\to 0$ and $\mu \to 0$ even in samples free from \hee3 impurities. [\hee3 suppresses superflow and syringe \cite{Hallock,Beamish,Beamish_2016}]. Observing such a suppression without suppressing superflow would be a "smoking gun" for the superclimb mechanism \cite{sclimb}  and for the emergence of LL. 

As mentioned in Ref.\cite{PRB2015}, the current experiments \cite{Hallock,Beamish} and also \cite{Beamish_2016} are likely to be in the regime of large $\mu$, that is, in the dislocation rough state  induced by the bias where $\kappa=\kappa_1=\kappa_g$, Eq.(\ref{Kap_0}), even at $T=0$. In this case, the equilibrium regime, when the flow has essentially stopped and the resulting pressure increase is measured vs the imposed one, can provide valuable information. 
  The number of the injected (or removed) atoms should be $ \delta N \sim \mu/G$, where $G$ is the shear modulus of ideal crystal, in the limit of non-interacting dislocations.
However, dislocations interact through elastic forces. This can be described by global compression energy (density) increase $\Delta E \approx \Delta P^2/(2K_{el})$, where $\Delta P$ is the global pressure change and $K_{el}$ is the compression modulus. Thus, the response $\Delta   P$ on imposed $\mu$ is $\Delta P \approx \mu K_{el} /G$. The  modulus $K_{el}$ has a contribution from the full shear modulus $G_r$ \cite{Landau_el}, which in its turn is known to exhibit  softening as $T$ increases, Ref.\cite{Beamish_el}, due to the glide of basal plane dislocations. Thus,
$\Delta P$ should show the response which is {\it increasing} with {\it decreasing } $T$: $\Delta P \sim (G_r/G) \mu$.  In other words,
$\Delta P(T)/\mu$ measured by syringe should show the same type of $T$-dependence observed in measurements of the plasticity \cite{Beamish_el}. 

\noindent {\it Conclusions}.
We have introduced the J-current type model (\ref{S_J}) describing tilted superclimbing dislocation. According to the elementary scaling analysis such a dislocation should exhibit non-LL behavior. In contrast,  Monte Carlo simulations reveal the emergence of the LL as temperature is lowered and the system size exceeds certain scale determined by the line tension $G$ (bare shear modulus). This scale is characterized by high power independent of the bare superfluid stiffness. The emerging LL shows also the BKT transition into insulating state.The LL behavior can be destroyed by {\it macroscopically} small external bias by chemical potential. As a result, the giant isochoric compressibility  can be reinstated even at $T=0$.

\noindent{\it Acknowledgements}. We thanks Boris Svistunov for fruitful discussions. This work was supported by  the NSF grant PHY1314469.

\end{document}